\documentclass[aps,prl,reprint,preprintnumbers,showpacs,showkeys,superscriptaddress]{revtex4-1}
\usepackage{amsmath,amssymb,bm,mathrsfs,graphicx}
\usepackage{srcltx}
\usepackage[colorlinks=true,citecolor=blue,linkcolor=blue]{hyperref}

\begin{document}
\title{Absence of Quantum Time Crystals}

\author{Haruki Watanabe}
\email{hwatanabe@berkeley.edu}
\affiliation{Department of Physics, University of California, Berkeley, California 94720, USA}

\author{Masaki Oshikawa}
\email{oshikawa@issp.u-tokyo.ac.jp}
\affiliation{Institute for Solid State Physics, University of Tokyo, Kashiwa 277-8581, Japan} 

\begin{abstract}
In analogy with crystalline solids around us, Wilczek recently proposed
 the idea of ``time crystals'' as phases that spontaneously break the continuous time translation into a discrete subgroup. The proposal stimulated further studies and vigorous debates whether it can be realized in a physical system. However, a precise definition of the time crystal is needed to resolve the issue. Here we first present a definition of time crystals based on the time-dependent correlation functions of the order parameter. We then prove a no-go theorem that rules out the possibility of time crystals defined as such, in the ground state or in the canonical ensemble of a general Hamiltonian, which consists of not-too-long-range interactions.
\end{abstract}
\maketitle

Recently, Wilczek proposed a fascinating new concept of time
crystals, which spontaneously break the continuous time translation
symmetry, in analogy with ordinary crystals which break the continuous
spatial translation symmetry~\cite{Wilczek1,BrunoComment1,Wilczek2}.
Li {\it et al.} soon followed with a concrete proposal for
an experimental realization and observation of a (space-)time crystal,
using trapped ions in a ring threaded by an Aharonov-Bohm
flux~\cite{TongcangLi,BrunoComment2,TongcangLi2}.
While the proposal of time crystals was quite bold,
it is, on the other hand, rather natural
from the viewpoint of relativity:
since we live in the Lorentz invariant space-time,
why don't we have time crystals if there are ordinary crystals
with a long-range order in spatial directions?

However, the very existence, even as a matter of principle,
of time crystals is rather controversial.
For example, Bruno~\cite{Bruno} and Nozi\`{e}res~\cite{Nozieres} discussed some difficulties in realizing time crystals.
However, since their arguments were not fully
general, several new realizations
of time crystals, which avoid these no-go arguments,
were proposed~\cite{Wilczek3,Nitta}.

In fact, a part of the confusion can be attributed to
the lack of a precise mathematical definition of time crystals.
Here, we first propose a definition of time crystals
in the equilibrium, which is a natural generalization of that
of ordinary crystals and can be formulated precisely also
for time crystals.
We then prove generally the absence of time crystals
defined as such, in the equilibrium with respect to
an arbitrary Hamiltonian which consists of not-too-long-range
interactions.
We present two theorems:
one applies only to the ground state,
and the other applies to the equilibrium with an arbitrary
temperature.

Naively, time crystals would be defined in terms of the expectation
value $\langle\hat{O}(t)\rangle$ of an observable $\hat{O}(t)$.  If
$\langle\hat{O}(t)\rangle$ exhibits a periodic time dependence, the
system may be regarded as a time crystal.  However, the very definition
of eigenstates $\hat{H}|n\rangle=E_n|n\rangle$ immediately implies that
the expectation value of any Heisenberg operator
$\hat{O}(t)\equiv e^{i\hat{H}t}\hat{O}(0)e^{-i\hat{H}t}$
in the Gibbs equilibrium ensemble
is time independent.  To see this, recall that the expectation value
$\langle\hat{X}\rangle$ is defined as
$\langle\hat{X}\rangle\equiv\langle0|\hat{X}|0\rangle$ at zero
temperature $T=0$ and
$\langle\hat{X}\rangle\equiv\text{tr}(\hat{X}e^{-\beta\hat{H}})/Z=\sum_n\langle
n|\hat{X}|n\rangle e^{-\beta E_n}/Z$ at a finite temperature
$T=\beta^{-1}>0$, where $|0\rangle$ is the ground state and $Z\equiv
\text{tr}[e^{-\beta\hat{H}}]$ is the partition function.  Clearly,
$\langle n|\hat{O}(t)|n\rangle$ is time independent since two factors of
$e^{\pm iE_nt}$ cancel against each other and hence
$\langle\hat{O}(t)\rangle$ is time independent.

Yet it is too early to reject the idea of time crystals just from this
observation, since a similar argument would preclude ordinary (spatial)
crystals.  One might naively define crystals from a spatially modulating
expectation value of the density operator $\hat{\rho}(\vec{x}) =
e^{-i\hat{\vec{P}}\cdot\vec{x}}\hat{\rho}(\vec{0})
e^{i\hat{\vec{P}}\cdot\vec{x}}$.  The unique ground state of the
Hamiltonian in a finite box is nevertheless symmetric and hence
$\hat{\vec{P}}|0\rangle=0$, implying that
$\langle\hat{\rho}(\vec{x})\rangle$ is constant over space at
$T=0$. Likewise, at a finite temperature,
$\langle\hat{\rho}(\vec{x})\rangle\equiv\text{tr}(e^{-i\hat{\vec{P}}\cdot\vec{x}}\hat{\rho}(\vec{0})e^{i\hat{\vec{P}}\cdot\vec{x}}e^{-\beta\hat{H}})/Z$
cannot depend on position since $\hat{\vec{P}}$ and $\hat{H}$ commute.
More generally, the equilibrium expectation value
of any order parameter vanishes in a finite-size system,
since the Gibbs ensemble is always symmetric.
This, of course, does not rule out the possibility of
spontaneous symmetry breaking.

A convenient and frequently used prescription to
detect a spontaneous symmetry breaking is 
to apply a symmetry-breaking field.
For example, in the case of antiferromagnets on a cubic lattice, we apply a
staggered magnetic field $h_s(\vec{R})=h\cos(\vec{Q}\cdot\vec{R})$
[$\vec{Q}\equiv(\pi/a)(1,\ldots,1)$] by adding a term
$-\sum_{\vec{R}}h_s(\vec{R})\hat{s}_{\vec{R}}^z$ to the Hamiltonian,
where $\vec{R}$'s are lattice sites and $\hat{s}_{\vec{R}}^z$ is the
spin on the site $\vec{R}$.  One computes the expectation value of the
macroscopic order parameter, which is the staggered magnetization in the
case of an antiferromagnet, under the symmetry breaking field and then
take the limit $V\rightarrow\infty$ and $h\rightarrow0$ in this order.
The non-vanishing expectation value of the macroscopic order
parameter, in this order of limits, is often regarded as
a definition of spontaneous symmetry breaking (SSB).
In the case of crystals, we apply a potential $v(\vec{x})=h
\sum_{\vec{G}}v_{\vec{G}}\cos(\vec{G}\cdot\vec{x})$ with a periodic {\it
position} dependence.  Here, $\vec{G}$'s are the reciprocal lattice of
the postulated crystalline order.

This prescription is quite useful but unfortunately is not
straightforwardly applicable to
time crystals.  The symmetry-breaking field for time crystals has to
have a periodic \emph{time} dependence.  In the presence of such a
field, the ``energy'' becomes ambiguous and is defined only modulo the
frequency of the periodic field, making it difficult to select states
or to take statistical ensembles based on energy eigenvalues.  
Therefore an alternative definition of time crystals
is called for, and indeed we will propose a definition of time crystals which
is applicable to very general Hamiltonians.

\paragraph{Time-dependent long-range order.}
---In order to circumvent the problem in defining time crystals using
a time-dependent symmetry-breaking field,
here we define time crystals based on the long-range behavior of
correlation functions.
In fact, all conventional symmetry breakings can be defined in terms of
correlation functions,
without introducing any symmetry-breaking field. 
That is, we say the system has
a \emph{long-range order} (LRO)
if the equal-time correlation function of the
local order parameter $\hat{\phi}(\vec{x},t)$ satisfies
\begin{equation}
\lim_{V\rightarrow\infty}\langle \hat{\phi}(\vec{x},0)\hat{\phi}(\vec{x}',0)\rangle\rightarrow \sigma^2\neq0\label{LRO}
\end{equation}
for $|\vec{x}-\vec{x}'|$ much greater than any microscopic scales.  
One can equivalently use the integrated order parameter 
$\hat{\Phi}\equiv \int_V \mathrm{d}^dx\, \hat{\phi}(\vec{x},0)$,
for which the long-range order is defined as
$\lim_{V\rightarrow\infty}\langle\hat{\Phi}^2\rangle/V^2=\sigma^2\neq0$.
For example, in the case of the quantum transverse Ising model
$\hat{H} = - \sum_{\langle \vec{r}, \vec{r}' \rangle}
\sigma^z_{\vec{r}} \sigma^z_{\vec{r}'}
- \Gamma \sum_{\vec{r}} \sigma^x_{\vec{r}}$,
the local order parameter $\hat{\phi}(\vec{r},t)$ is identified with
$\sigma^z_{\vec{r}}$ or its coarse-graining.
It has been proven quite generally that the LRO $\sigma\neq0$ 
guarantees the corresponding SSB, namely a non-vanishing expectation
value of the order parameter in the limit of the zero symmetry-breaking
field taken after the limit $V \to \infty$~\cite{KHL,KomaTasakiCMP}.
While the reverse is not proved in general, it is expected to hold in many systems of interest.

A crystalline order can also be defined by the correlation function.
Namely, if the long-range correlation approaches to a periodic
function
\begin{equation}
\lim_{V\rightarrow\infty}\langle
\hat{\phi}(\vec{x},0)\hat{\phi}(\vec{x}',0)\rangle\rightarrow f(\vec{x}-\vec{x}')\label{LRO2}
\end{equation}
for sufficiently large $|\vec{x}-\vec{x}'|$, the system exhibits a
spontaneous crystalline order~\cite{XiaoGang}.  Equivalently, $\lim_{V\rightarrow\infty}\langle \hat{\Phi}_{\vec{G}}\hat{\Phi}_{-\vec{G}}\rangle/V^2=f_{\vec{G}}\neq0$ signals a density wave order with wavevector $\vec{G}$, where 
$\hat{\Phi}_{\vec{G}}=\int_V\mathrm{d}^dx\,\hat{\phi}(\vec{x},0)e^{-i\vec{G}\cdot\vec{x}}$.
Note again that
$\langle\hat{\phi}(\vec{x},0)\rangle$ itself is a constant over
space in the Gibbs ensemble, which is symmetric.
For instance, we set
$\hat{\phi}=\hat{\rho}$ for ordinary crystals and
$\hat{\phi}=\hat{s}_{\alpha}$ for spin-density
waves.
In terms of the LRO, one can therefore characterize crystals using only the
symmetric ground state or ensemble, which itself does not have a finite
expectation value of the order parameter~\cite{Griffiths}. 

\begin{figure}
\begin{center}
\includegraphics[width=\columnwidth]{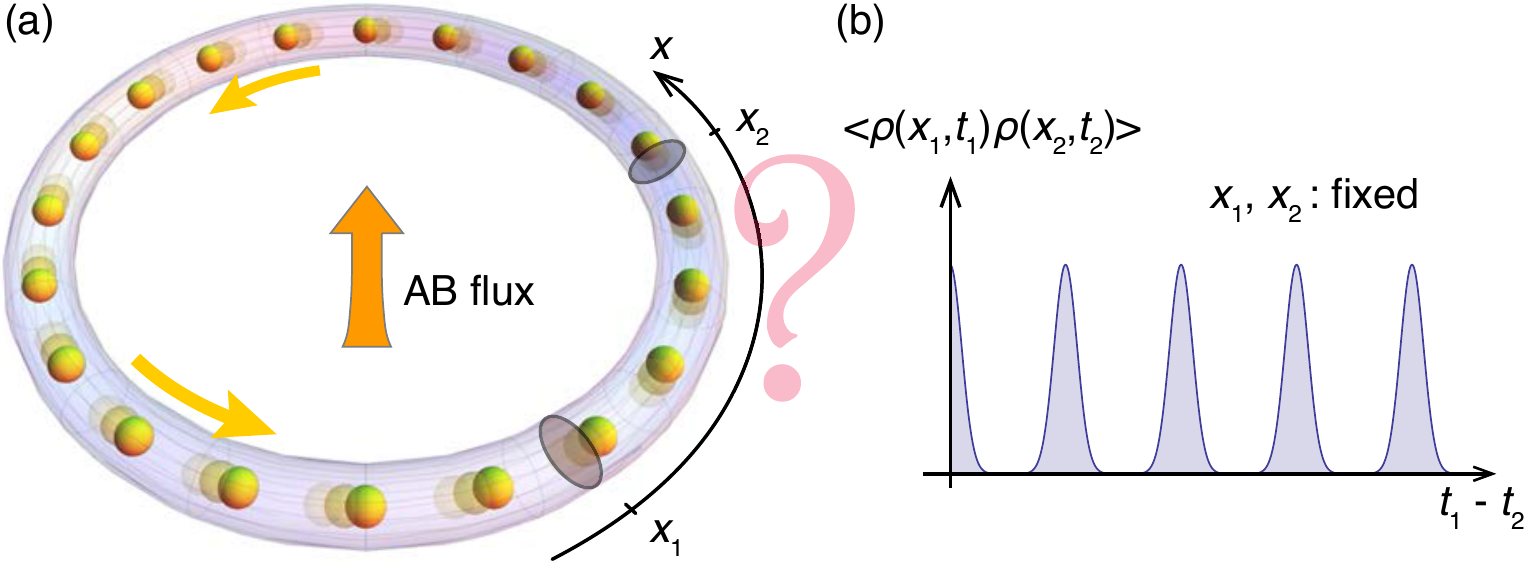}
\caption{ Time-dependent correlation. 
(a) Wigner crystal of ions in a ring threaded by an Aharonov-Bohm
flux, as proposed in Ref.~\cite{TongcangLi}
as a possible realization of a time crystal.
(b) Illustration of the time dependent correlation function, should
the time crystal is indeed realized as a spontaneous rotation
of the density wave (crystal) in the ground state, as proposed.
The density-density correlation function between $(x_1,t_1)$ and
$(x_2,t_2)$ must exhibit an oscillatory behavior
as a function of $t_1-t_2$ for fixed $x_1$ and $x_2$.}
\label{fig1}
\end{center}
\end{figure}

Let us now define time crystals, in an analogous manner to the
characterization of ordinary crystals in terms of the spatial LRO.
Generalizing Eqs.~\eqref{LRO} and \eqref{LRO2}, we could say the
system is a time crystal if the correlation function
$\lim_{V\rightarrow\infty}\langle\hat{\phi}(\vec{x},t)\hat{\phi}(0,0)\rangle\rightarrow f(t)$
is non-vanishing for large enough $|\vec{x}|$ and
exhibits a nontrivial periodic oscillation in time $t$ (i.e., is not
just a constant over time).
In terms of the integrated
order parameter
defined above, the condition reads
\begin{eqnarray}
\lim_{V\rightarrow\infty}\langle e^{i\hat{H}t}\hat{\Phi}e^{-i\hat{H}t}\hat{\Phi}\rangle/V^2=f(t).
\label{LRO3}
\end{eqnarray}

When $f$ is a periodic function of both space and time,  
we call it a space-time crystal, in which case we have
\begin{equation}
\lim_{V\rightarrow\infty}\langle e^{i\hat{H}t}\hat{\Phi}_{\vec{G}}e^{-i\hat{H}t}\hat{\Phi}_{-\vec{G}}\rangle/V^2=f_{\vec{G}}(t),\label{LRO4}
\end{equation}
where $f_{\vec{G}}(t)$ is the Fourier component of $f(t,\vec{x})$.
For example,
Li {\it et al.}~\cite{TongcangLi}
investigated a Wigner crystal in a ring threaded by a
Aharonov-Bohm flux and predicted its spontaneous rotation,
which would be a realization of space-time crystal.
If this were indeed the case,
the density at $(x_1,t_1)$ and $(x_2,t_2)$ would be correlated
as illustrated in Fig.~\ref{fig1}.

One might think that we could define time crystals based on the time
dependence of equal-{\it position} correlation functions.  
Should we adopt this definition, however, rather trivial
systems would qualify as time crystals.
For example, consider a two-level system $\hat{H}=-\Omega_0\sigma_z/2$ at $T=0$
and set $\hat{\phi}(t)\equiv\sigma_x(t)=e^{i\hat{H}t}\sigma_x
e^{-i\hat{H}t}=\sigma_x\cos\Omega_0 t+\sigma_y\sin\Omega_0t$.  The
correlation function $\langle 0|\hat{\phi}(t)\hat{\phi}(0)|0\rangle$ of
the ground state $|0\rangle=(1,0)^T$ exhibits a periodic time dependence
$e^{-i\Omega_0 t}$.
The same applies to the equal-position correlation function
in independent two-level systems spread over the space.
Clearly we do not want to classify such a trivial, uncorrelated
system as a time crystal.
``Crystal'' should be reserved for systems exhibit
correlated, coherent behaviors, which are captured by
long-distance correlation functions, be it an ordinary crystal
or a time crystal.

\paragraph{Absence of long-range time order at $T=0$.}
---We now prove quite generally that time crystals defined above are not
possible in ground states.
More precisely, we show
\begin{equation}
\frac{1}{V^2}\left|\langle
0|\hat{A}e^{-i(\hat{H}-E_0)t}\hat{B}|0\rangle-\langle
0|\hat{A}\hat{B}|0\rangle\right|\leq C\frac{t}{V}, \label{statement}
\end{equation}
where $E_0$ is the ground-state energy.
Equation~\eqref{statement} holds for any Hermitian operators
$\hat{A}=\int_V\mathrm{d}^dx\,\hat{a}(\vec{x})$ and
$\hat{B}=\int_V\mathrm{d}^dx\,\hat{b}(\vec{x})$, where
$\hat{a}(\vec{x})$ and $\hat{b}(\vec{x})$ are local operators that act
only near $\vec{x}$.  The constant $C$ may depend on $\hat{A}$,
$\hat{B}$, and $\hat{H}$ but not on $t$ or $V$.  Once we prove
Eq.~\eqref{statement}, we can immediately see that $f(t)$ in
Eq.~\eqref{LRO3} is time independent, by setting
$\hat{A}=\hat{B}=\hat{\Phi}(0)$ and taking the limit
$V\rightarrow\infty$ for $t=o(V)$. We can also apply Eq.~\eqref{statement} to space-time crystals characterized by $f(t,\vec{x})$.
Although $\hat{\Phi}_{\vec{G}}$ may not be Hermitian, one can always decompose it
to the sum of two Hermitian operators. Applying
Eq.~\eqref{statement} for each of them, one can see all
$f_{\vec{G}}(t)$'s, and hence $f(t,\vec{x})$, are time independent.

To show Eq.~\eqref{statement}, we use the trick to represent the change in
time by an integral:
\begin{eqnarray}
&&\left|\langle 0|\hat{A}e^{-i(\hat{H}-E_0)t}\hat{B}|0\rangle-\langle
0|\hat{A}\hat{B}|0\rangle\right|\notag\\
&&=\left|\int_0^t\mathrm{d}s\frac{\mathrm{d}}{\mathrm{d}s}\langle
0|\hat{A}e^{-i(\hat{H}-E_0)s}\hat{B}|0\rangle\right|\notag\\
&&\leq\int_0^t\mathrm{d}s\left|\langle
0|\hat{A}(\hat{H}-E_0)e^{-i(\hat{H}-E_0)s}\hat{B}|0\rangle\right|.
\label{proof1}
\end{eqnarray}
The integrand can be bounded by the Schwarz inequality as
\begin{eqnarray}
&&\left|\langle
0|\hat{A}(\hat{H}-E_0)^{1/2}e^{-i(\hat{H}-E_0)s}(\hat{H}-E_0)^{1/2}\hat{B}|0\rangle\right|\notag\\
&&\leq \sqrt{\langle 0|\hat{A}(\hat{H}-E_0)\hat{A}|0\rangle\langle
0|\hat{B}(\hat{H}-E_0)\hat{B}|0\rangle}\notag\\
&&=\frac{1}{2}\sqrt{\langle
0|[\hat{A},[\hat{H},\hat{A}]]|0\rangle\langle
0|[\hat{B},[\hat{H},\hat{B}]]|0\rangle}.  \label{proof2}
\end{eqnarray}
Each of $\hat{H}$, $\hat{A}$, and $\hat{B}$ involves a spatial
integration and introduces a factor of $V$, while each commutation
relation reduces a factor of $V$, assuming that the equal-time
commutation relation of any two operators $\hat{\phi}_1(\vec{x},t)$ and
$\hat{\phi}_2(\vec{x}',t)$ can be nonzero only near $\vec{x}=\vec{x}'$.
Hence, $\|[\hat{A},[\hat{H},\hat{A}]]\|$ is at most of the order of
$V^{3-2}=V$~\cite{Horsch,KomaTasakiJSP}.  The same is true for
$\|[\hat{B},[\hat{H},\hat{B}]]\|$.  Therefore, combining
Eqs.~\eqref{proof1} and \eqref{proof2}, we get the desired
Eq.~\eqref{statement}.

In this estimate of $\|[\hat{A},[\hat{H},\hat{A}]]\|$, we assumed the
locality of the Hamiltonian, i.e., $\hat{H}$ is an integral of 
Hamiltonian density $\hat{h}(\vec{x})$, which contain only
local terms.
It is easy to see that the same conclusion holds even when
there are interactions among distant points,
provided that the interaction decays exponentially as a
function of the distance.
One can further relax this assumption to power-law
decaying interactions $r^{-\alpha}$ ($\alpha>0$). When $0<\alpha<d$,
$\|[\hat{A},[\hat{H},\hat{A}]]\|$ can be order of $V^{2-(\alpha/d)}$ and
the right-hand side of Eq.~\eqref{statement} should be accordingly
modified to $CtV^{-(\alpha/d)}$, where $d$ is the spatial dimension of
the system.  When $\alpha\geq d$, Eq.~\eqref{statement} holds
without any change. In both cases, as long as $\alpha>0$, $f(t)$ remains
time independent in the limit $V\rightarrow\infty$ for a fixed finite
$t$.  

\paragraph{Absence of long-range time order at a finite $T$.}
---The argument presented above cannot directly be extended to excited
eigenstates $|n\rangle$, because $(\hat{H}-E_0)^{1/2}$ in
Eq.~\eqref{proof1} would then be replaced by $(\hat{H}-E_n)^{1/2}$
but the latter is not well defined.  Instead, here we employ the
Lieb-Robinson bound~\cite{LiebRobinson} to discuss finite temperatures.

The result of Lieb and Robinson is that~\cite{LiebRobinson}
\begin{equation}
\|[e^{i\hat{H}t}a(\vec{x})e^{-i\hat{H}t},b(\vec{y})]\|\leq
\text{min}\{C_1,C_2e^{-\mu (|\vec{x}-\vec{y}|-vt)}\},\label{LRbound}
\end{equation}
where constants $C_{1,2}$, $\mu$, and $v$ may depend on $\hat{a}$,
$\hat{b}$, and $\hat{H}$. This bound is valid only for a local
Hamiltonian. The physical meaning of Eq.~\eqref{LRbound} is that
there exists an upper bound on the velocity at which information can
propagate in quantum systems.

To prove the time independence of $f(t)$ in Eq.~\eqref{LRO3} and
$f_{\vec{G}}(t)$ in Eq.~\eqref{LRO4}, let us introduce a new
correlation function defined by the commutation relation
\begin{eqnarray}
g_{AB}(t)&\equiv&\big\langle\big[e^{i\hat{H}t}\hat{A}e^{-i\hat{H}t},\hat{B}\big]\big\rangle/V^2\notag\\
&=&\int_V\mathrm{d}^dx\,\big\langle\big[e^{i\hat{H}t}\hat{a}(\vec{x})e^{-i\hat{H}t},\hat{b}(\vec{0})\big]\big\rangle/V.
\end{eqnarray}
The Lieb-Robinson bound \eqref{LRbound} tells us that
$|g_{AB}(t)|\leq [C_3+C_4(vt)^d]/V$
for some constants $C_{3,4}$ that do not depend on $t$ or $V$.  Hence,
as long as $t=o(V^{1/d})$, $|g_{AB}(t)|\rightarrow0$ as
$V\rightarrow\infty$.

On the other hand, we have $g_{AB}(t)=f_{AB}(t)-f_{BA}(-t)$, where $f_{AB}(t)\equiv\big\langle e^{i\hat{H}t}\hat{A}e^{-i\hat{H}t}\hat{B}\big\rangle/V^2$.
By inserting the complete set $1=\sum_n|n\rangle\langle n|$, it can be
readily shown that
\begin{eqnarray}
f_{AB}(t)&=&\int_{-\infty}^{\infty}\mathrm{d}\omega\,\rho_{AB}(\omega)e^{-i\omega
t},\\ g_{AB}(t)&=&\int_{-\infty}^{\infty}\mathrm{d}\omega\,(1-e^{-\beta
\omega})\rho_{AB}(\omega)e^{-i\omega t},\label{relationfg}
\end{eqnarray}
where $\rho_{AB}(\omega)$ is defined by
\begin{equation}
\sum_{n,m}\frac{\langle m|\hat{A}|n\rangle\langle n|\hat{B}|m\rangle
e^{-\beta E_m}}{ZV^2}\delta(\omega-E_n+E_m).
\end{equation}
Since $\lim_{V\rightarrow\infty}g_{AB}(t)=0$ for any given real
value of $t$, $(1-e^{-\beta
\omega})\lim_{V\rightarrow\infty}\rho_{AB}(\omega)=0$.
Combined by the sum rule $\int d\omega \; \rho_{AB}(\omega) = f_{AB}(0)
= \langle \hat{A}\hat{B} \rangle/V^2$,
we find
\begin{equation}
\lim_{V\rightarrow\infty}\rho_{AB}(\omega)=\delta(\omega)\lim_{V\rightarrow\infty}\langle\hat{A}\hat{B}\rangle/V^2.
\end{equation}
Therefore, $f_{AB}(t)$, as a function of finite $t$
in the thermodynamic limit $V\to \infty$, 
can be at most a finite constant that does not
depend on time.
Thus (space-)time crystals do not exist at a finite temperature either.

\paragraph{Grand-canonical ensemble.}
---Let us discuss systems with variable number of particles.
The equilibrium of those systems can be described by
a grand-canonical ensemble.
It is given by the
Boltzmann-Gibbs distribution with respect to
$\hat{\cal H} = \hat{H} - \mu \hat{N}$,
where $\mu$ is the chemical potential determined by
the property of the particle reservoir.
Namely, the expectation value of an observable $\hat{X}$ is
given by $\langle\hat{X}\rangle_{\mu}\equiv \text{tr}
(\hat{X}e^{-\beta \hat{\cal H}})/Z_\mu$, 
where $Z_\mu\equiv\text{tr}\,e^{-\beta \hat{\cal H}}$.
Although the statistical weight is given in terms of $\hat{\cal H}$,
the time evolution of the Heisenberg operator $\hat{\Psi}(t)$
is still defined by $\hat{H}$, {\it i.e.},
$\hat{\Psi}(t)\equiv e^{i\hat{H}t}\hat{\Psi}(0)e^{-i\hat{H}t}$.
This mismatch can produce some trivial time dependence
as we shall see now.
If we define $\hat{\Psi}_\mu(t)\equiv
e^{i(\hat{H}-\mu\hat{N})t}\hat{\Psi}(0)e^{-i(\hat{H}-\mu \hat{N})t}$ and
assume $[\hat{N},\hat{\Psi}(0)]=-q\hat{\Psi}(0)$ with $q$ a real number
that represents the U(1) charge of $\Psi$, then $\hat{\Psi}(t)=\hat{\Psi}_\mu(t)e^{-iq\mu t}$.
Therefore, even if
$f_\mu=\lim_{V\rightarrow\infty}[\langle\hat{\Psi}_\mu(t)\hat{\Psi}_\mu^\dagger(0)\rangle_\mu/V^2]$
is time independent as we proved above,
$f(t)=\lim_{V\rightarrow\infty}[\langle\hat{\Psi}(t)\hat{\Psi}^\dagger(0)\rangle_\mu/V^2]$
has a trivial time dependence $f(t)=f_\mu e^{-iq\mu t}$.  This is
consistent with the well-known fact that the order parameter of a
Bose-Einstein condensate has the trivial time dependence~\cite{Pethick}
as $\langle\hat{\psi}(\vec{x},t)\rangle=\psi_0 e^{-i\mu t}$.
This type of time dependence has been discussed~\cite{Nicolis} also in
the context of time crystals~\cite{Castillo,Thies}.
However, Volovik pointed out that this kind of
time dependence cannot be measured as long as the particle number is
exactly conserved~\cite{Volovik}.  Indeed, the overall phase of
condensate cannot be measured unless one couples the condensate to
another one.
We will discuss this phenomenon in the following section.

\paragraph{Spontaneous oscillation of non-equilibrium states.}
---In order to extract the time-dependence of the condensate order
parameter, the system has to be attached to another system
to allow change of the number of particles.
As a simplest setup,
we may prepare two condensates
with different chemical potentials $\mu_1,\mu_2$ and measure their
time-dependent interference pattern $\propto e^{-i(\mu_1-\mu_2) t}$
in terms of the current between the condensates,
or equivalently the change of the number of particles in each
condensate.
This is nothing but the ac Josephson effect.
In fact, in Ref.~\cite{Wilczek3}, a proposal of
time crystal based on this effect was made.

However, in order to observe the ac Josephson effect,
the initial state simply must not be in the equilibrium.
In order to see this, it is helpful to use the mapping
of the ac Josephson effect in two coupled condensates
to a quantum spin in a magnetic field.
For simplicity, let us consider condensates of bosons without
any internal degree of freedom, 
and suppose there is only one single-particle state in
each condensate.
Then the system can be described by the two set of
bosonic annihilation/creation operators,
$a, a^\dagger$ and $b,b^\dagger$.
The effective Hamiltonian of the system, in the limit of zero coupling
between the two condensates, is given as
  $H = \mu_1 a^\dagger a + \mu_2 b^\dagger b
    = (\mu_1 - \mu_2) \frac{a^\dagger a - b^\dagger b}{2}
    + \frac{\mu_1 + \mu_2}{2} N$,
where $N = a^\dagger a + b^\dagger b$ is the total number
of particles in the coupled system.
Let us assume that the coupling to the outside environment
is negligible in the timescale we are interested.
$N$ is then exactly conserved and can be regarded as a constant.
As a consequence, the second
term in the Hamiltonian proportional to $N$ can be ignored.

With $N$ being exactly conserved, this system of coupled condensates
can be mapped to a quantum spin model by identifying the bosons
as Schwinger bosons.
The Hamiltonian now reads
  $H =  B S^z +\mbox{const.}$,
where $B = \mu_1 - \mu_2$ and $S^z$ is the $z$-component of the
quantum spin with the spin quantum number $S= (N-1)/2$.
Similarly, the current operator between the two condensates is given by
$J \propto -i (a^\dagger b  - b^\dagger a) = 2 S^y$.
The ac Josephson effect, in the quantum spin language, is
just a Larmor precession about the magnetic field.
The oscillatory behavior of the current in the ac Josephson
effect just corresponds to the oscillation of the
exctation value of $S^y$ in the Larmor precession.

In order to observe the Larmor precession, the initial
state must have a non-vanishing expectation value
of the transverse component ($S^x$ or $S^y$).
This excludes the ground state, in which the spin is fully
polarized along the magnetic field in $z$ direction,
as well as thermodynamic equilibrium at arbitrary temperature.
In Ref.~\cite{Wilczek3} it was argued that, by taking
the limit of weak coupling, the dissipation can be
made arbitrarily small.
While this is certainly true, the lack of dissipation does not
mean that the system is in an equilibrium, as it is clear
by considering the spin Larmor precession in a magnetic field.
Our result, which is valid for equilibrium,
of course does not exclude such spontaneous
oscillations of \emph{non-equilibrium} quantum states.
The latter, however, are well known and should not be called
time crystals without a further justification.

\paragraph{Discussion.}---In this Letter, we proposed a definition of time crystals and proved their absence in the equilibrium.  The present result brings back the question:
why there is no time crystal, even though there surely exist
crystals with a spatial long-range order?
We should recall that Lorentz invariance does not mean the complete
equivalence between space and time: the time direction is still
distinguished by the different sign of the metric.
This leads to a fundamental difference in the spectrum:
while the eigenvalues of the Hamiltonian
(the generator of translation in time direction)
is bounded from below, the eigenvalues of the momentum
(the generator of translation in a spatial direction)
is unbounded.
Moreover, the equilibrium is determined by the Hamiltonian
and the system is generally not Lorentz invariant in
the thermodynamic equilibrium.
Therefore, as far as the equilibrium as defined in
standard statistical mechanics is concerned,
it is not surprising to
find a fundamental difference between space and time.

\begin{acknowledgments}
HW wishes to thank Ashvin Vishwanath for his
useful and important suggestions.
The authors are very grateful to Hal Tasaki and Tohru Koma for helping us improving the presentation. 
In particular, the proof for $T=0$ presented in this paper is an improved version due to Tohru Koma, after discussion with Hal Tasaki.
The work of MO was supported by the ``Topological Quantum Phenomena''
(No. 25103706) Grant-in Aid for
Scientific Research on Innovative Areas from the Ministry of Education,
Culture, Sports, Science and Technology (MEXT) of Japan.
HW appreciates the financial support from the Honjo
International Scholarship Foundation.
\end{acknowledgments}

\end{document}